\documentclass[journal,12pt,onecolumn,draftclsnofoot,]{IEEEtran}

\ifCLASSINFOpdf

\else

\fi

\usepackage{url}

\usepackage{mathrsfs}
\usepackage{amssymb}
\usepackage{amsmath}

\usepackage{amsthm}
\usepackage{cite}
\usepackage{epsfig}
\usepackage{graphicx}
\usepackage{subcaption}
\usepackage{hyperref}
\usepackage{algorithm,algorithmicx,algpseudocode}
\usepackage{url}
\usepackage{epstopdf}
\usepackage{color}
\usepackage{multirow}
\usepackage{mathtools}

\newcommand{\op}[1]{\ensuremath{\mathcal{#1}}}
\newcommand{\dif}{{\mathrm{d}}}

\newcommand{\vc}[1]{\ensuremath{\boldsymbol{#1}}}

\newtheorem{thm}{Theorem}[section]
\newtheorem{lem}[thm]{Lemma}

\theoremstyle{definition}

\newtheorem{cond}{Condition}[section]
\theoremstyle{remark}

\begin{document}

\title{Consistency Analysis for  the Doubly Stochastic Dirichlet Process}

\author{Xing Sun,~\IEEEmembership{Student Member, IEEE,}
         Nelson H.C.~Yung,~\IEEEmembership{Senior Member, IEEE,}
         Edmund Y.~Lam,~\IEEEmembership{Fellow, IEEE,}
        and Hayden K.-H.~So,~\IEEEmembership{Senior Member, IEEE}
\thanks{X. Sun, E. Lam, and H. So are with the Department
of Electrical and Electronic Engineering, The University of Hong Kong, Pokfulam, Hong Kong
(email: hso@eee.hku.hk ).}
\thanks{N. Yung is with Yung \& Partners Limited, Hong Kong.}
\thanks{Manuscript received May, 2016}}

\markboth{Technical Report}%
{Shell \MakeLowercase{\textit{et al.}}: Bare Demo of IEEEtran.cls for Journals}

\maketitle

\begin{abstract}
This technical report  proves components consistency for the  Doubly Stochastic Dirichlet Process~\cite{xing2016DSDP} with exponential convergence of posterior probability. We also present the fundamental properties for DSDP as well as inference algorithms. Simulation toy experiment and real-world experiment results for single and multi cluster also support the consistency proof.
This report is also a support document for the paper ``Computationally Efficient Hyperspectral Data Learning Based on the Doubly Stochastic Dirichlet Process''~\cite{xing2016DSDP}.  
\end{abstract}


\IEEEpeerreviewmaketitle

\section{The asymptotic mass function and exchangeable partitions}
\label{section_asymptotic}
The probability of data partitions is important in mixture modeling~\cite{ferguson1973bayesian}.
Let $\op{M}$ be the unordered partition of $n$ observations, 
then the probability mass function~\cite{antoniak1974mixtures} of $\op{M}$ follows.
\begin{lem}
\label{thm_SDSDP}
Let $D$ denote a  DSDP-MM with a Marked SGP prior thinning function $q'(\vc{\theta}_k, n_k) = q(\vc{\theta}_k)\times Q(n_k)$. 
The probability mass function of the unordered data partition $\op{M}$ follows
\begin{equation}
\label{formula_SDSDP}
P(\op{M}) = \dfrac{({\alpha^{*}})^K \Gamma(\alpha^*)}{\Gamma(\alpha^* + n) } \cdot \prod\limits_{k=1}^K {\Big\{\Gamma(n_{k}) \cdot Q(n_{k}) \Big\}},
\end{equation}
where $n_{k}$ is the number of $k^{th}$ data partition with the topic ${\vc{\theta}}_k$, and the number of land-cover classes is denoted as $K$. 
\end{lem}
Let $\op{M}^r = (\op{M}_1^r,\ldots,\op{M}_{|\op{M}|}^r)$ be the ordered derivation of random partition $\op{M}$, and the order is uniformly sampled from all $|\op{M}|!$ possible choices~\cite{miller2015mixture}. 
Let random vector $\op{N}=(\op{N}_1,\ldots,\op{N}_{|\op{M}|})$ be the size vector of $\op{M}^r$ and $\vc{n} = (n_1,\ldots,n_K)$, then the ordered partition probability follows
\begin{equation}\label{S_SDSDP}
\begin{split}
 &P(\op{N}  = \vc{n}) = \sum\limits_{\op{M}^r:\op{N}(\op{M}^r)=\vc{n} }{\dfrac{P(\op{M})}{|\op{M}|!}}\\
 &= \dfrac{n!}{K!}  \dfrac{B(n)^K }{{\Gamma (a_0)^K}} \dfrac{{b_0}^{K\cdot a_0}n^n}{ (b_0 +n)^{K\cdot a_0 +n} }
\dfrac{{\alpha^{*}}^K \Gamma(\alpha^*)}{\Gamma(\alpha^* + n) } \prod\limits_{k=1}^K {\dfrac{  {\Gamma{(a_0 + n_k)} }}{n_k \cdot n_k !}}.
\end{split}
\end{equation}
To obtain asymptotic partition probability,  the asymptotic marked function is derived first.
\begin{lem}
\label{thm_Asym}
For any cluster amount $K \in \{ 1, 2, \ldots  \}$ and any finite hyper-parameter $a_0>0$, if the mixture weight of the partition would always be nonzero $\frac{n_{\vc{\theta}}}{n} >0$, as the data amount grows $n\to \infty$, the asymptotic marked function follows
\begin{equation}
\label{formula_Asym}
Q(n_{\vc{\theta}})\sim  \dfrac{{(n_{\vc{\theta}}+1)}^{a_0-1}}{n^{a_0}}.
\end{equation}
\end{lem}

As $n\to\infty$, the ordered partition probability follows
\begin{equation}\label{S_SDSDP_limit}
 P(\op{N} =\vc{n}) \sim  \dfrac{{(\alpha^{*} b_0^{a_0})}^K }{e\cdot n^{(\alpha^* + K\cdot a_0)}\cdot{\Gamma (a_0)^K}}\cdot \dfrac {n!}{K!}\cdot\prod\limits_{k=1}^K {n_k^{(a_0-2)}}.
\end{equation}
When the number of partitions $|\op{M}|$ is given,
the conditional partition size probability $P(\op{N} =\vc{n} \mid|\op{M}|=K) $ of DSDP Mixture ${P}_{DS}$, the Mixture of Finite Mixture (MFM) ${P}_{FM}$~\cite{miller2015mixture} and the DP Mixture ${P}_{DP}$ follows
\begin{equation}\label{SDSDP_T}
{P}_{DS} \propto \prod\limits_{k=1}^K {n_k^{(a_0-2)}},{P}_{FM} \propto\prod\limits_{k=1}^K {n_k^{(\gamma-1)}}, {P}_{DP} \propto \prod\limits_{k=1}^K {n_k^{-1}},
\end{equation}
where $\gamma$ is the hyper-parameter for MFM. 
When $a_0 = \gamma+1$, the size probability ${P}_{DS}$ of DSDP is the same as ${P}_{FM}$ of the MFM. The size  probability ${P}_{DS}$ is equal to ${P}_{DP}$ of the DP when $a_0 = 1$. 
Overall, Eq.~\ref{SDSDP_T} illustrates that all of these three models are shaped as a symmetric $K$-dimensional Dirichlet distribution, and the partition is exchangeable.

Discussions above assume that the weight of each partition is nonzero, as the correct mixture weights are nonzero. {Here, a general situation in HSI identification is discussed, where some partition amounts (HSI data amounts of some land-cover classes) may be quite small.}

\begin{lem}
\label{thm_Asym2}
For any partition amount $n_{\vc{\theta}}\geq 1$, any cluster amount $K \in \{ 1, 2, \ldots  \}$and any finite hyper-parameter $a_0>0$, as $n\to \infty$, asymptotic marked function follows

\begin{equation}
\label{formula_Asym3}
Q(n_{\vc{\theta}})\sim \dfrac{\Gamma(a_0 + {n_{\vc{\theta}}})}{   \Gamma(a_0) \Gamma({n_{\vc{\theta}}} ) } \cdot   \dfrac{{n_{\vc{\theta}}} ^{-1}}{{(n+b_0)}^{a_0}}.
\end{equation}
\end{lem}
This lemma is more general for any partition amount $n_{\vc{\theta}}$.
\begin{thm}
For any sampled partition size ${n_k}\in \mathbb{Z}^{+}, k=1,\ldots,K$, when number of clusters is given with $|\op{M}| = K$, the conditional partition size probability of DSDP-MM follows
 \begin{equation}\label{SDSDP_T2}
P\Big(\op{N} =\vc{n} \bigr\rvert|\op{M}|=K\Big) \propto \prod\limits_{k=1}^K {n_k^{-1}\dfrac{\Gamma(a_0 + {n_k})}{   \Gamma(a_0) \Gamma({n_k} ) }}.
\end{equation}
\end{thm}
\section{Consistency theorem proof}
\label{proof_con}
First, the following proof is based on the background knowledge of the conjugate prior in~\cite{diaconis1979conjugate}.
Suppose that the observation $\vc{X}$ is sampled from an exponential family distribution ${p_{\vc{\theta}} }({\vc{x}_j})$,
then ${T_1}\left( {{\vc{x}_j}} \right)$, ${h}_0(\vc{\theta})$ and ${A_1}\left( \vc{\theta}  \right) $ are the sufficient statistic,
the underlying measure and the log normalization, respectively. 
The natural parameter ${\eta _0} = \left\langle {{\eta _1},{\eta _2}} \right\rangle $ is the hyper-parameter of the base distribution ${H_0}({\vc{\theta}} |{\eta _0}) $.

Second, to derive probabilities \textit{\textbf{p}} and \textit{\textbf{q}},
a marginal probability of DSDP-MM follows
 \begin{equation}
  \label{marginal_probability}
  \begin{split}
  {m_q}({{\vc{\vec x}}_{{{\rm A}_i}}}) 
  =  {\int_\Theta  {H_0{(\vc{\theta})}\sigma \left( Y({\vc{\theta}} ) \right)\prod\limits_{j \in {{\rm A}_i}} {{p}({{\vc{x}}_j}|\vc{\theta})}\dif{\vc{\theta}} } }
    = {{\prod\limits_{j \in {{\rm A}_i}} {{p_0}({{\vc{x}}_j})} }}\cdot{\int_\Theta{e^{ {\eta_{{{\rm A}_{i}}}} f_{\rm{A}_i}( \vc{\theta})}\sigma \left( Y({\vc{\theta}} ) \right)\dif\vc{\theta}}} ,
 \end{split} 
 \end{equation}
 where ${{\eta_{{{\rm A}_{i}}}}} = {\left( {{\eta _2}+ \left| {{{\rm A}_i}} \right|} \right)}$ is the natural parameter for the data partition ${{{\rm A}_{i}}}$.
 The function follows 
 $ f_{\rm{A}_i}( \vc{\theta}) = {{ {\vc{\mu}_{{\rm A}_i}^T{\vc{\theta}}  { - {{A}_1}\left( {\vc{\theta}}  \right)}}  } }+{\eta^{-1}_{{\rm A}_i}}{\log({h}_0(\vc{\theta}))}$, 
where  $\vc{\mu}_{{\rm A}_i} = {\eta^{-1}_{{\rm A}_i}} {{ ({ \sum_{j \in {{\rm A}_i}} {{T_1}( {{{\vc{x}}_j}})} }+\eta_1) }}$.

Subsequently, 
observations in the newly separated cluster have the same original topic,
as the Split-Merge MCMC is applied.
The extra partition is the subset of a original data partition  $ {\rm{A}}_{K+1} \subset {\rm{B}}_i, i \leq K$.
The data partition probability in Theorems V.1 and V.2 in~\cite{xing2016DSDP} is derived as
\begin{align}\label{pp}
 {{P\left( {{T_n} = K,{{\vc{x}}_{1:n}}} \right)}}
 =  &\sum\limits_{{\rm B} \in {{\rm A}_{K}}(n)} {p({\rm B}){{\prod\limits_{i = 1}^K{{m_q}({{\vec{\vc{x}}}_{{{\rm B}_i}}})} }}}  \cr
 =& \dfrac{{c_{\Gamma}}{c_{Q}}N_{\{\rm B \to \rm A\}}}{|\op{R}_{\rm{A}}(\op{M},\op{C}_{\varphi})|}
 \times\sum\limits_{{\rm A} \in {{\rm A}_{K+1}}(n)} {{c_{\varphi}}({\rm{A}})  p({\rm A}){{\prod\limits_{i = 1}^{K+1} {{m_q}({{\vec{\vc{x}}}_{{{\rm A}_i}}})} }}} \cr
 =& C_{\Phi} \cdot{{p\left( {{T_n} = K+1,{{\vc{x}}_{1:n}}} \right)}},
\end{align}
where ${c_{\Gamma}}{c_{Q}}$ are mass and marked function ratios defined in Eq.~\ref{cc} in the Section~\ref{app_a}.
The $\op{R}_{\rm{A}}(\op{M},\op{C}_{\varphi})$ is the set of all possible new partitions ${\rm{A}}_{K+1}$ such that its parent partition ${\rm{A}}_l $ satisfies ${{m_q}({{\vec{\vc{x}}}_{{{\rm A}_l}}})} = c_{\varphi}{{m_q}({{\vec{\vc{x}}}_{{{\rm A}_{K+1}}}})}{{m_q}({{\vec{\vc{x}}}_{{{\rm A}_l}\smallsetminus{{\rm A}_{K+1}} }})}$ for all ${{\rm A} \in {{\rm A}_{K+1}}(n)}$, which subjects to  $c_{\varphi}({\rm{A}})\in \op{C}_{\varphi}$. 

Lastly, based on partition probability, 
four cases are provided to prove Theorems V.1 and V.2 in~\cite{xing2016DSDP}.
\begin{enumerate}
\item
First, if the sampled data partition amount $K$ is correct in the previous iteration $(t-1)$,
then it is impossible that any original data partition ${\rm{B}}_{l}$ would split into two partitions ${\rm{A}}_{l}$, ${\rm{A}}_{K+1}$ at current iteration $(t)$ when
${{p( {{T_n^{(t-1)}} = K} )}} \geq c  { n}^{s_0}{{p( {{T_n^{(t)}} = K+1} )}} $.
This equation is satisfied with the conclusion $P_n(C_{\Phi} \geq c  { n}^{s_0} ) = 1$, which is shown in ~\ref{finite_split_app}.

\item
The second case shows the situation that the partition amount $K+1$ has been correctly sampled in the previous iteration $(t-1)$. 
Then it is impossible that two partitions ${\rm{A}}_{l}$, ${\rm{A}}_{K+1}$ would merge when
${{p( {{T_n^{(t)}} = K} )}} \leq c  { n}^{-v_0}{{p( {{T_n^{(t-1)}} = K+1} )}} $.
See the conclusion $P_n(C_{\Phi} \leq  c{{n}^{-v_0}} ) =1$ in ~\ref{infite_merge_app} for details.

\item
Third, if the sampled partition amount $K+1$ in the previous iteration $(t-1)$ is larger than actual amount $K$,
there exists a nonzero possibility that two partitions ${\rm{A}}_{l}$, ${\rm{A}}_{K+1}$ would merge at this iteration $(t)$ when
${{p( {{T_n^{(t)}} = K} )}} \geq c  \cdot{{p( {{T_n^{(t-1)}} = K+1} )}} $.
The conclusion $P_n(C_{\Phi} \geq c   ) = 1$ in ~\ref{finite_merge_app} explains details.

\item
Lastly, if the situation that the sampled partition amount $K$ in the previous iteration $(t-1)$ is smaller than the actual amount $K+1$,
there exists the possibility that a partition ${\rm{B}}_{l}$ would split into two partitions ${\rm{A}}_{l}$, ${\rm{A}}_{K+1}$ at this iteration $(t)$ when
${{p( {{T_n^{(t-1)}} = K} )}} \leq c  \cdot{{p( {{T_n^{(t)}} = K+1} )}} $.
It is satisfied with the conclusion $P_n(C_{\Phi} \leq  c ) =1$  in ~\ref{infite_split_app}.
\end{enumerate}

\section{Posterior Probability Ratios and Conditions}
\label{app_a}
In Eq.~\ref{pp}, 
coefficient $C_{\Phi}$ contains five major elements.

The first two elements are the mass coefficient $ {c_{\Gamma}}$,
and the marked coefficient ${c_{Q}}$,  
which are derived with the Eq.~\ref{formula_Asym3}, 
\begin{equation}
  \label{cc}
 {c_{\Gamma}}{c_{Q}} = \dfrac{{|{{\vec{\vc{x}}}_{{{\rm A}_l}}}|}^{\frac{3}{2}}{|{{\vec{\vc{x}}}_{{{\rm A}_{K+1}}}}|}^{\frac{3}{2}}}{{|{{\vec{\vc{x}}}_{{{\rm B}_l}}}|}^{\frac{3}{2}}{(n+b_0)}^{-a_0}} \dfrac{{\Gamma(a_0)}{\Gamma(a_0 +|{{\vec{\vc{x}}}_{{{\rm B}_l}}}|)}}{{\Gamma(a_0 +|{{\vec{\vc{x}}}_{{{\rm A}_l}}}|)}{\Gamma(a_0 +|{{\vec{\vc{x}}}_{{{\rm A}_{K+1}}}}|)}}.
\end{equation}

The third factor $N_{\{\rm B \to \rm A\}}$ is the number of possible partitions ${\rm{B}} \in {{\rm{A}}_{K}}(n)$ that splitting into  ${\rm A} \in {{\rm A}_{K+1}}(n)$. 
In the Algorithm~\ref{algorithm3}, ${\rm A}_{K+1}$ is the subset of one original partition, 
where at most $K$ possibilities. 
Therefore, $N_{\{\rm B \to \rm A\}} \in \{0,1, \ldots, K\}$.

The fourth element is the marginal probability ratio $c_{\varphi}$,
for any partition ${{\rm A} \in {{\rm A}_{K+1}}(n)}$, 
the ratio follows
\begin{subequations}
\label{ratio_m}
\begin{alignat}{3}
{c_\varphi } (\rm A)=& \dfrac{{{{m _q}({{{\vc {\vec{x}}}}_{{{\rm A}_l}}})} }{{{m _q}({{{\vc {\vec{x}}}}_{{{\rm A}_{K+1}}}})} }} {{{m _q}({{{\vc {\vec x}}}_{{{\rm B}_{l}}}})}}\\
{=} & c_Y\dfrac{ {\int_\Theta{e^{ {\eta_{{{\rm A}_{l}}}} f_{\rm{A}_l}( \vc{\theta})}\dif\vc{\theta}}}\cdot{\int_\Theta{e^{ {\eta_{{{\rm A}_{K+1}}}} f_{\rm{A}_{K+1}}( \vc{\theta})}\dif\vc{\theta}}}} {{\int_\Theta{e^{ {\eta_{{{\rm B}_{l}}}} f_{\rm{B}_l}( \vc{\theta})}\dif \vc{\theta}}}}  \\
{=}& c_{Y} c_{L} \omega_r { {\prod\limits_{i=l,K+1}e^{-{\eta_{{{\rm A}_{i}}}} ||f_{\rm{A}_i}( \vc{\theta}_{\rm{A}_i}^*)-f_{\rm{A}_i}( \vc{\theta}_{\rm{B}_l}^*)||}}},
\end{alignat}
\end{subequations}
where $ \omega_r = \frac{\sqrt{{\eta_{{{\rm B}_{l}}}}}}{\sqrt{{{{\eta_{{{\rm A}_{l}}}}}{{\eta_{{{\rm A}_{K+1}}}}} }}}$, and $c_Y, c_L$ are finite constants.
Eq.~\ref{ratio_m}~(b) is the ratio for normalized constant. 
Coefficient $c_Y>0$ is finite when SGP function $\sigma \left( Y({\vc{\theta}} ) \right)$ is removed,
which is shown in Section~\ref{sub:gaussian_process_intensity}.
In Eq.~\ref{ratio_m}~(c), Laplace's method~\cite{azevedo1994laplace} has been applied to approximate the normalization constant.
The function $ f({\vc{\theta}})$ is the posterior probability of the parameter ${\vc{\theta}}$, 
and the extreme value location $\vc{\theta}^*$ is the topic of any data set $\vc{x}_{s}$.
For the Laplace's method, the function $f({\vc{\theta}}) $ is second-order differentiable and its second order derivative at $\vc{\theta}^*$ follows $f''({\vc{\theta}^*})  <0$, due to that $f''({\vc{\theta}^*})    = -  \text{var}(\vc{x}_{s})$~\cite{davison2003statistical}.

The last factor is the possible partition amount $|\op{R}_{\rm{A}}(\op{M},\op{C}_{\varphi})|$, which is discussed in Section~\ref{Nodistance} and Section~\ref{Distance}.

\subsection{The finite mixture weight condition}
  \begin{cond}
If there exists a finite DP Mixture weight prior $\pi _0 (\vc{\theta}) < \infty, \forall {\vc{\theta}}  \in \Theta$, 
the topic parameter or data partition amount is finite.
As the whole data amount grows,
for any sample data partition ${{{\vc{X}}}_{{{\rm B}}}}$, 
its amount $|{{\rm B}}|$ follows that 
\label{Finite prior2}
\begin{equation}
\lim\limits_{n\to\infty}P(\dfrac{|{{\rm B}}|}{n} > 0) = 1.
\end{equation}
\end{cond}

\subsection{Data partitions with the same topic parameter}
\label{Nodistance}
Based on the Slutsky's theorem~\cite{bradley1962asymptotic}, if observations are sampled from distributions with the same topic $\vc{\theta}_0$, as the data amount $|{\rm A}| \to \infty$, 
 inferred topic parameters ${{\vc{\theta}}_{\rm{A}}^*}$ of any sampled data partition ${\rm A}$ will approach the real topic $\vc{\theta}_0$, i.e.
\begin{equation}
  \label{asy_exponitial}
{{\vc{\theta}}_{\rm{A}}^*} =  A_1^{'-1}\left(\dfrac{1}{|{\rm A}|}\sum\limits_{i\in{ {\rm{A}}}}{T_1(\vc{X}_i)}\right) \xrightarrow{P}  \vc{\theta}_0,
\end{equation}
where $A_1^{'-1}$ is the inverse function of the derivative of the log normalization. 
See~\cite{davison2003statistical} for detailed asymptotic properties. 
Hence, for any two data partitions $\rm{A}$ and $\rm{B}$, whose data are sampled from the same topic $\vc{\theta}_0$ with Condition~\ref{Finite prior2}. 
The function $f_{\rm A}$ in Eq.~\ref{ratio_m} can be derived under this setting. 
As data amount grows, there exists a constant $c>0$, which subjects to
\begin{equation}
\label{cc1}
\begin{split}
  P\big\{|| {{\vc{\theta}}_{\rm{A}}^*}  - {{\vc{\theta}}_{\rm{B}}^*}   ||<\frac{c}{\min\{{|{\rm A}|}, {|{\rm B}|}\}} \big\}& = 1,\\
  P\big\{|| f_{\rm{A}}( \vc{\theta}_{\rm{A}}^*) - f_{\rm{A}}( \vc{\theta}_{\rm{B}}^*)   || < \frac{c}{|{\rm{B}}|} \big\} &= 1,
\end{split}
\end{equation}
where $\vc{\theta}_{\rm{A}}^*, \vc{\theta}_{\rm{B}}^*$ are inferred topics for data partition ${\rm A}$ and ${\rm B}$.

Let $w_{{{\rm A}}} = a_0 +|{{\vec{\vc{x}}}_{{{\rm A}}}}|-1$ for any data partition ${\rm{A}}$.
If data partitions are sampled from the same topic, 
two situations are discussed with the amount of data partition ${w_{{{\rm A}_{K+1}}}}$ is finite: 
finite split with a correctly inferred SGP prior, 
and finite merge with a incorrectly inferred SGP prior.
The possible partition number follows $|\op{R}_{\rm{A}}(\op{M},\op{C}_{\varphi})|\leq ({w_{{{\rm A}_{K+1}}}})! = c_g$, and it is finite at these situations when ${w_{{{\rm A}_{K+1}}}}$ is finite.
Hence, the coefficients $C_{\Phi} $ can be derived, with
\begin{equation}
\label{seting1}
\lim\limits_{n\to\infty} C_{\Phi}=   C_1\cdot{\bar{c}_{\varphi}}{{n}^{(w_{{{\rm{A}}_{K+1}}}+1)}},
\end{equation}
where $C_1 = \frac{ {{|{{\vec{\vc{x}}}_{{{\rm A}_{K+1}}}}|}^{\frac{3}{2} }}\Gamma(a_0)}{c_g \Gamma(w_{{{\rm A}_{K+1}}}+1)} >0$ is finite. 
The mean ratio $\bar{c}_{\varphi}$ is the weighting average for all data partitions ${{\rm A} \in {{\rm A}_{K+1}}(n)}$ in Eq.~\ref{pp}.
This derivation is based on substituting Eq.~\ref{cc} 
and the Gamma function ratio $\lim\limits_{n\to\infty} \frac{{\Gamma(a_0 +|{{\vec{\vc{x}}}_{{{\rm B}_l}}}|)}}{{\Gamma(a_0 +|{{\vec{\vc{x}}}_{{{\rm A}_l}}}|)}}  =c{ n}^{|{{{\rm A}_{K+1}}}|}$ into Eq.~\ref{pp}, in which $c>0$ is a finite constant.

Substituting Eq.~\ref{cc1} into Eq.~\ref{ratio_m},
then for any partition $\rm{A} \in {{\rm A}_{K+1}}(n)$, 
the ratio ${c_{\varphi}}(\rm{A})$ can be derived from Eq.~\ref{seting1}, where
\begin{equation}
\label{finite_equ}
\begin{split}
\lim\limits_{n\to\infty}&P_n(C_{\Phi} \geq c  { n}^{s_0} )\\
&=\lim\limits_{n\to\infty}P_n \left({c_{\varphi}}({\rm{A}}) > c \cdot\frac{{n}^{(s_0-w_{{{\rm{A}}_{K+1}}})} }{|{\rm{B}}|} \right)  = 1.
\end{split}
\end{equation}

The value of the variable $s_0$ determines which situation Eq.~\ref{finite_equ} is suitable.
\subsubsection{Finite Split}
\label{finite_split_app}
 $s_0 = w_{{{\rm A}_{K+1}}}>0$.
Partition ${\rm{B}}_{l}$ is impossible to split into two partitions ${\rm{A}}_{l}$, ${\rm{A}}_{K+1}$ at this iteration $(t)$, when sampling in the previous iteration is correct.
  
\subsubsection{Finite Merge}
\label{finite_merge_app}
$s_0 = 0$.
The probability that merges two sampled partitions ${\rm A}_{l}$ and ${\rm A}_{K+1}$ in the previous iteration step to compose a correct new partition ${\rm{B}}_{l}$ is larger than zero.

Discussion for infinite split with the correct SGP prior and infinite merge with the incorrect SGP prior are quite similar.

\subsection{Data partitions with different topic parameters}
\label{Distance}
For any two data partitions ${\rm{A}}, {\rm{B}}$, 
whose observations are sampled from two different topics $\vc{\theta}_{a0}, \vc{\theta}_{b0}$ with $ ||{\vc{\theta}_{a0}-\vc{\theta}_{b0}}||> 0$. 
As the data amounts of these two partitions grow, 
there exists a finite constant $c>0$, which subjects to
\begin{equation}
\label{cc3}
\begin{split}
    P\Big(|| {{\vc{\theta}}_{\rm{A}}^*}  - {{\vc{\theta}}_{\rm{B}}^*}   ||> c \Big) &= 1\cr
     P\Big(|| f_{\rm{A}}( \vc{\theta}_{\rm{A}}^*) - f_{\rm{A}}( \vc{\theta}_{\rm{B}}^*)   || > c \Big) &= 1,
\end{split}
\end{equation}
where $\vc{\theta}_{\rm{A}}^*, \vc{\theta}_{\rm{B}}^*$  are estimated topics (maximum posterior parameters) for data partitions ${\rm A}$ and  ${\rm B}$, respectively.

In this situation, we discuss partitions, which have finite and nonzero mixture weight.  
The amount ratios follow that
$\lim\limits_{n\to\infty}\frac{w_{{{\rm A}_{K+1}}}}{w_{{{\rm B}_l}}} =c_1>0$, 
$1 - c_1 = \lim\limits_{n\to\infty}\frac{w_{{{\rm A}_l}}}{w_{{{\rm B}_l}}} >0$.
 The possible partition number follows $|\op{R}_{\rm{B}}(\op{M},\op{C}_{\varphi})|>1$ at these situations.
The coefficients $C_{\Phi}$ can be derived with the asymptotic expression, 
${{\Gamma(w_{{{\rm B}_l}}+1)}}\sim \sqrt{2\pi w_{{{\rm B}_l}}} {\big(\frac{ w_{{{\rm B}_l}}}{e}  \big)}^{ w_{{{\rm B}_l}}}$.
\begin{equation}
\begin{split}
\label{CC2}
 \lim\limits_{n\to\infty}C_{\Phi} &= \dfrac{C_1\cdot{\bar{c}_{\varphi}}{{|{{\vec{\vc{x}}}_{{{\rm B}_l}}}|}^{(a _0+ \frac{3}{2})}}\sqrt{2\pi w_{{{\rm B}_l}}} {\big(\frac{ w_{{{\rm B}_l}}}{e}  \big)}^{ w_{{{\rm B}_l}}}}{2\pi \sqrt{w_{{{\rm A}_l}}w_{{{\rm A}_{K+1}}}} {\big(\frac{ w_{{{\rm A}_l}}}{e}  \big)}^{ w_{{{\rm A}_l}}} {\big(\frac{ w_{{{\rm A}_{K+1}}}}{e}  \big)}^{ w_{{{\rm A}_{K+1}}}}}\cr
 & = C_2\cdot{\bar{c}_{\varphi}}{{|{{\vec{\vc{x}}}_{{{\rm B}_l}}}|}^{(a _0+1)}} \cdot {c_u}^{{ w_{{{\rm B}_l}}}},
\end{split}
\end{equation}
where  $C_1 = \Gamma(a_0) {{ (c_1-c_1^2)}}^{1.5}\frac{{{(n+b_0)}}^{a_0} }{ {w_{{{\rm B}_l}}^{a_0}}}$ is finite with Condition~\ref{Finite prior2},
and $C_2 =  \frac{e }{ \sqrt{2\pi (c_1-c_1^2) }}C_1 $. 
The mean ratio $\bar{c}_{\varphi}$ is the weighting average for all data partitions ${{\rm B} \in {{\rm B}_{K}}(n)}$ in Eq.~\ref{pp}.
This derivation is based on substituting Eq.~\ref{cc} 
and the Gamma function ratio:
$ {\{ {({c_1})}^{-c_1}{(1-{c_1})}^{-1+c_1}\}}^{ w_{{{\rm B}_l}}}$ into Eq.~\ref{pp}. 
Since $c_1\in (0,1)$, 
the constant  $c_u \in (1,2)$.
Substitute Eq.~\ref{cc3} into Eq.~\ref{ratio_m},
then for any partition $\rm{B} \in {{\rm B}_{K}}(n)$, 
the ratio ${c_{\varphi}}(\rm{B})$ can be further derived from Eq.~\ref{CC2}, where
\begin{equation}
\label{infinite_result_merge}
\begin{split}
\lim\limits_{n\to\infty}&P_n(C_{\Phi}\leq  c{{n}^{-v_0}} ) \cr
&=\lim\limits_{n\to\infty} P_n({c_{\varphi}}(\rm{B})< {2}^{-{ w_{{{\rm B}_{l}}}}}{{|{{\vec{\vc{x}}}_{{{\rm B}_{l}}}}|}^{-(v_0+a_0+1)}})=1.
\end{split}
\end{equation}

Two situations of the variable $v_0$ in Eq.~\ref{infinite_result_merge} require further comments.
\subsubsection{Infinite Split}
\label{infite_split_app}
$v_0 = 0$.
The probability that split an incorrect partition ${\rm{B}}_{l}$ into two partitions ${\rm A}_{l}$ and ${\rm A}_{K+1}$ is larger than zero.

\subsubsection{Infinite Merge}
\label{infite_merge_app}
$v_0 > 0$.
The probability that merge two correctly sampled partitions ${\rm A}_{l}$ and ${\rm A}_{K+1}$ in the previous iteration step to compose a new partition ${\rm{B}}_{l}$ is zero.

Discussions for finite merge with the correct SGP prior and finite split with the incorrect SGP prior are not provided,
as these two situations do not exist based on Condition~\ref{Finite prior2}.

\section{Gaussian Process Intensity} 
\label{sub:gaussian_process_intensity}
To maximize the likelihood in Eq. 9 in~\cite{xing2016DSDP}, 
the likelihood $\ln p( {{{{\vc{Y}}}},|\vc{\vec{\theta}},M,K} ) $ can be set to zero. 
Then we can obtain 
${\vc Y_{K + M}} = {{\vc \Sigma }_{(K + M )\times (K + M)}}{\left[ 
   {\sigma \left( { - {{Y}_{1}}} \right)}  
    \ldots   
   { - \sigma \left( {{{ Y}_{K + M}}} \right)} 
 \right]}^T$. 
Here we introduce two variables used for sampling the SGP Cox Process~\cite{adams2009tractable}. 
The variable ${a_{ins}}$ indicates the probability to add a new latent variable with the corresponding GP function $Y_{new}$ 
and the variable ${a_{del}}$ is the probability to delete an existing  latent variable with $Y_{old}$. 
Increase and delete ratios will be balanced at ${a_{ins}} = {a_{del}}$ when the convergence is achieved.
When we safely set the proposal probability $b(\cdot,\cdot) = 0.5$,
the Condition~\ref{Finite prior2}is satisfied for dataset ${\vc{x}}_{1:n}$, 
and the upper bound parameter ${\alpha ^*}$ is finite, 
we can conclude amounts of the topic parameters and latent variables follow
\begin{equation}
\mathop {\lim }\limits_{n \to \infty } \mathop {\sup }\limits_{c \in \left( {0,1} \right)} P\left( \left( K + M \right) < c \alpha ^* \right) = 1. 
\end{equation}
For any latent topic parameter $\vc{\theta}_k \in {\vc{\theta}}_{1:K}$ , the corresponding GP function $Y_k$ can be derived as
\begin{equation}
    {Y_k} =  \sum\limits_{i = 1}^K {\frac{{{e^{\left( { - {Y_i}} \right)}}k\left( {{{\vc{\theta}}_k},{{\vc{\theta}}_i}} \right)}}{{1 + {e^{\left( { - {Y_i}} \right)}}}}}  - \sum\limits_{i = K + 1}^{K + M} {\frac{{k\left( {{{\vc{\theta}}_k},{{\vc{\theta}}_i}} \right)}}{{1 + {e^{\left( { - {Y_i}} \right)}}}}}. 
\end{equation}
The GP function $Y_m$ of latent variable is similar. If $K$ and $M$ are finite, sampled GP functions follow
$  |Y_i| < \infty$,  $i\in \{1,\ldots,K+M\}$.
Therefore, the SGP function $\sigma ( {{Y_i}} ) $ follows
\begin{equation}
\mathop {\lim }\limits_{n \to \infty } P\left( {\sigma {\left( {{Y_i}} \right) }\in {\left(0,1\right)} }\right) = 1.
\end{equation}
Hence, for any data partition ${\rm A}_{i}$, there always exists a finite constant $c>0$, which subject to
\begin{equation}
  \dfrac{{\int_\Theta{e^{ {\eta_{{{\rm A}_{i}}}} f_{\rm{A}_i}( \vc{\theta})}\dif\vc{\theta}}}}{{\int_\Theta{e^{ {\eta_{{{\rm A}_{i}}}} f_{\rm{A}_i}( \vc{\theta})}\sigma \left( Y({\vc{\theta}} ) \right)\dif\vc{\theta}}}} = c.
\end{equation}

\section{Algorithms}
Here, we present two algorithms: Assignment Sampling for each HSI pixel ${\vc{x}}_i$ for DSDP-MM and Split-Merge MCMC Algorithm for DSDP-MM, which are shown in Algorithm~\ref{algorithm2} and Algorithm~\ref{algorithm3}, respectively.

\begin{algorithm}[t]
\caption{Assignment Sampling for each HSI pixel ${\vc{x}}_i$ for DSDP-MM with  the Marked SGP prior}
\label{algorithm2}
\begin{algorithmic}[1]
  \State {\bfseries Input:} {Likelihood $\ell ( {{\vc{x}}_i|\vc{\theta}_{k,i}} )$, data amount $n_{- i,k} $ and GP functions  ${{Y}_{1:K+M}^{\left( t \right)}}$}
  \State {\bfseries Output:} {Assignment sample $ \vec{z}_{i}^{(t + 1)}$}
   \For {$k=1:K$}
      \State Calculate the thinning function: {$q( {\vc{\theta}}_k, n_{{\vc{\theta}}_k}) \propto  n_{ - i,k}\cdot\sigma(Y_k) \cdot\dfrac{Q(n_{ - i,k}+1)}{Q(n_{ - i,k})}$} 
\EndFor
   \For {$k=K+1:K+M$}
      \State Calculate the thinning function: {$q({\vc{\theta}}_k, n_{{\vc{\theta}}_k}) \propto \dfrac{\alpha ^{*}}{M} \cdot \sigma(Y_k)\cdot {Q(1)} $}
\EndFor
   \State Sample assignment $ \vec{z}_{i}^{(t + 1)}$ via Eq. 11 in~\cite{xing2016DSDP}\;
  \end{algorithmic}
\end{algorithm}

\begin{algorithm}[t]
\caption{Split-Merge MCMC Algorithm for DSDP-MM with the Marked SGP prior (Split case)}
\label{algorithm3}
\begin{algorithmic}[1]
  \State {Random sample two data $x_i$ and $x_j$}  \Comment{Stage 1}
  \State {Let sampled distinct observation $x_i$ and $x_j$ crate two new cluster $\op{S}_{1}^{(0)},\op{S}_{2}^{(0)}$ respectively}  
  \State {Let $\op{S}_{\vc{c}}$ be the set of data which are belonged to $z_i$ or $z_j$ excluding  $x_i$ and $x_j$ in state $\vc{c}$}
  \State{let $t \leftarrow 0$} \Comment{Stage 2}
  \State{Randomly permute $\op{S}_{\vc{c}}$ with the random order function $\tau(\cdot)$}
  \If {$z_i=z_j=k$} \Comment{Split case}
  \State{${T}(\vc{c} \to \vc{c}_{split})   \leftarrow 1, \quad\quad  T(\vc{c}_{split} \to \vc{c})  \leftarrow 1 $}
    \For{random order $m \in \tau \left( 1 \right),...,\tau \left( |\op{S}_{\vc{c}}| \right)$}
  \State {Assignment sampling $z_m^r \sim  p ( {z_m |\op{S}_{1}^{(t)},\op{S}_{2}^{(t)}}  )$}  
  \State{$T(\vc{c} \to \vc{c}_{split})  \leftarrow T(\vc{c} \to \vc{c}_{split})p(z_m=z_m^r| \op{S}_{1}^{(t)},\op{S}_{2}^{(t)})  \prod\limits_{l=1,2}\dfrac{T\big(\vc{\theta}_{\op{S}_{l}^{(t)}} \to \vc{\theta}_{\op{S}_{l}^{(t-1)}}\big)}
  {T\big(\vc{\theta}_{\op{S}_{l}^{(t-1)}} \to \vc{\theta}_{\op{S}_{l}^{(t)}}\big) }$}
  \EndFor
\State {Calculate the acceptance ratio $\op{A}\leftarrow\dfrac{PG\ell(\vc{c}_{split})Q(|\vc{c}_{split}|)T( \vc{c}_{split}\to \vc{c} )}{PG\ell(\vc{c})Q(|\vc{c}|)T(\vc{c} \to \vc{c}_{split}) }$} 
\State{$t \leftarrow t+1$}
\Else{}
 \\\ldots\ldots \Comment{Merge case}
\EndIf
   \State Sample $u \sim \text{Unif}(0,1)$, if $u<\op{A}$ , accept the move; otherwise, reject it.\;  
  \end{algorithmic}
\end{algorithm}

\section{Experiment}
\label{Exp}
\subsection{Single Cluster Simulation} 

\begin{figure*}[!htp]
\begin{center}
\includegraphics[width=0.7\linewidth]{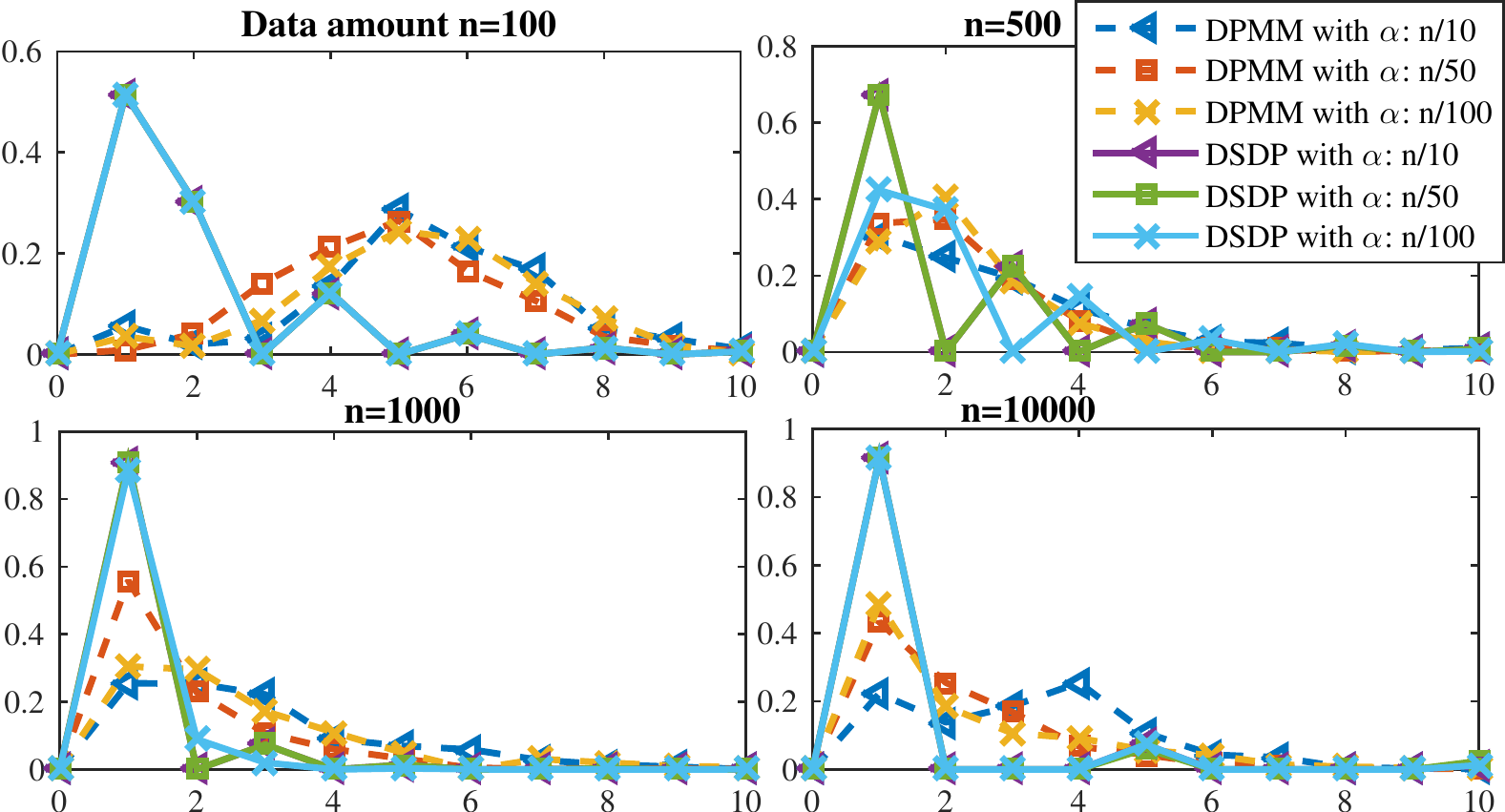}
\end{center}
\caption{Posterior probabilities of the cluster number with various
  initial concentration parameters $\alpha^*$. Clustering is applied
  on a single simulation dataset by DPMM and DSDP-MM.}
\label{1cluster_lamda}
\vskip -0.1in
\end{figure*}

The density sampling inference of DPMM may achieve consistent when the data amount grows,
but the inconsistency problem still exists for the number of clusters.
Moreover, this inconsistency problem could be severe for the single cluster dataset.
For a robust nonparametric Bayesian mixture model,
it is important that the cluster number could be consistent with
different values of the initial concentration parameter and the data amount.
Therefore,
this simulation experiment is built based on the following points:
\begin{enumerate}
\item
For the number of clusters, we applied the single cluster simulation data, which is the most severe inconsistency problem in~\cite{miller2013inconsistency}. 
\item
For the initial hyper-parameter, three typical values have been used: $n/10, n/50$ and $n/100$. 
\item
For the data amount, we analyze the range from ${10}^2$ to ${10}^4$, 
which is commonly used. 
\end{enumerate}
From Fig.~\ref{1cluster_lamda}, we can see that DPMM has severe inconsistency problem for the single cluster data~\cite{miller2013inconsistency}, 
but the proposed DSDP-MM can obtain a consistent result even for this situation.
In this simulation, different initial concentration parameters $\alpha$ have been used.  
The $x$-axis in Fig.~\ref{1cluster_lamda} shows the number of clusters. 
The $y$-axis represents the frequency of the number of clusters occurring in the last $1000$ iterations after convergence.

First, we discuss experimental results with various data amounts, 
four panels represent four different values of the data amount from $100$ to $10000$.
When the data amount is small, such as $n=100$ and $n=500$, 
the inconsistency becomes serious,
especially for DPMM.
DPMM would wrongly infer the data partition and the posterior probability of the cluster number with any initial concentration parameter when the data amount follows $n=100$. 
Subsequently, we employ different initial concentration parameters on all simulation data.
When the data amount grows, such as $n=1000$ and $n=10000$ in bottom panels,
 posterior probabilities (dotted line) of DPMM become quite inconsistent for different $\alpha^*$.
In bottom panels, posterior probabilities (solid line) of DSDP-MM are consistent and closer to the ground truth.
In conclusion, DSDP-MM obtains a more consistent data partition result compared to DPMM.

\subsection{Galaxy Experiment}

\begin{figure*}[!htp]
\begin{subfigure}{.5\textwidth}
  \centering
  \includegraphics[width=.85\linewidth]{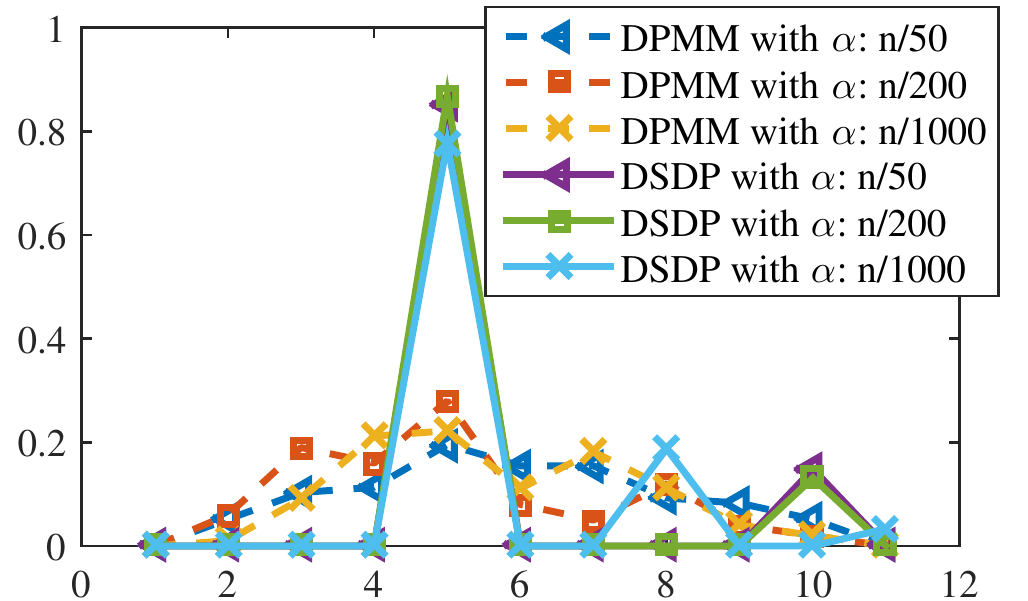}
  \caption{Posterior probabilities of the cluster number}
  \label{galaxy_num}
\end{subfigure}%
\begin{subfigure}{.5\textwidth}
  \centering
  \includegraphics[width=.85\linewidth]{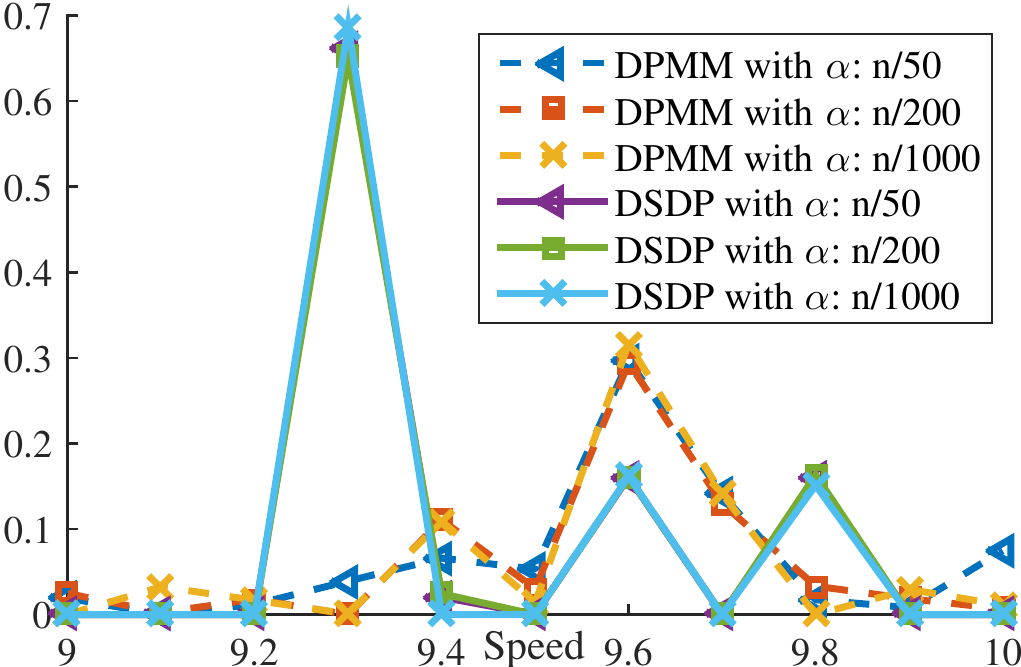}
  \caption{The density of the Galaxy data}
  \label{galaxy_density}
\end{subfigure}
\caption{Galaxy dataset clustering result for number of cluster and clusters density estimation}
\label{Galaxy_1}
\vskip -0.2in
\end{figure*}
Here, we further experiment on the real-world galaxy data~\footnote{This dataset can be downloaded at~\url{http://astrostatistics.psu.edu/datasets/Shapley_galaxy.html}},
which shows that DSDP-MM is adept in spatial clustering and estimating the density of the large-scale dataset. 
Shapley galaxy dataset with $4215$ galaxies in the Shapley Concentration regions~\cite{drinkwater2004large}, 
which is used to illustrate the extensibility of DSDP-MM to other related spatial data without rich feature.  
Right ascension (Coordinate in the sky similar to longitude on Earth), 
Declination (Coordinate in the sky similar to latitude on Earth) and speed comprise the three-dimension feature space.
This galaxy dataset is used to illustrate that DSDP-MM can achieve a consistent and robust result for large-scale dataset. 

Fig.~\ref{galaxy_num} also presents the posterior probabilities of the cluster number for DPMM and DSDP-MM, 
 as the data amount $n$ of this galaxy dataset is larger,
we set the initial concentration parameter $\alpha^*$ with $n/10$, $n/200$ and $n/1000$. 
The posterior probabilities of DSDP-MM at the reasonable data partition $K=5$ are all over $80\%$ for various $\alpha^*$, which are solid lines.
Dotted lines show that posterior probabilities of DPMM are closer to the uniformly distribution and the sampling would be inconsistent.
Therefore,
this figure demonstrates that results of DSDP-MM are much more consistent than DPMM for the number of clusters. 

Fig.~\ref{galaxy_density} analyzes the density of the topic parameters 
in speed feature dimension. 
Since some topic parameters of the speed are closer, 
we plot the density with discretized interval $0.1$.
DSDP-MM also has a more robust density as the initial concentration parameters $\alpha^*$ decrease from $\alpha = n/50$ to $\alpha = n/1000$. 
Densities of DPMM (dotted lines) differ greatly with different $\alpha^*$, 
especially for the speed from $9$ to $9.4$.

\begin{figure*}[!t]
\begin{center}
\includegraphics[width=0.82\linewidth]{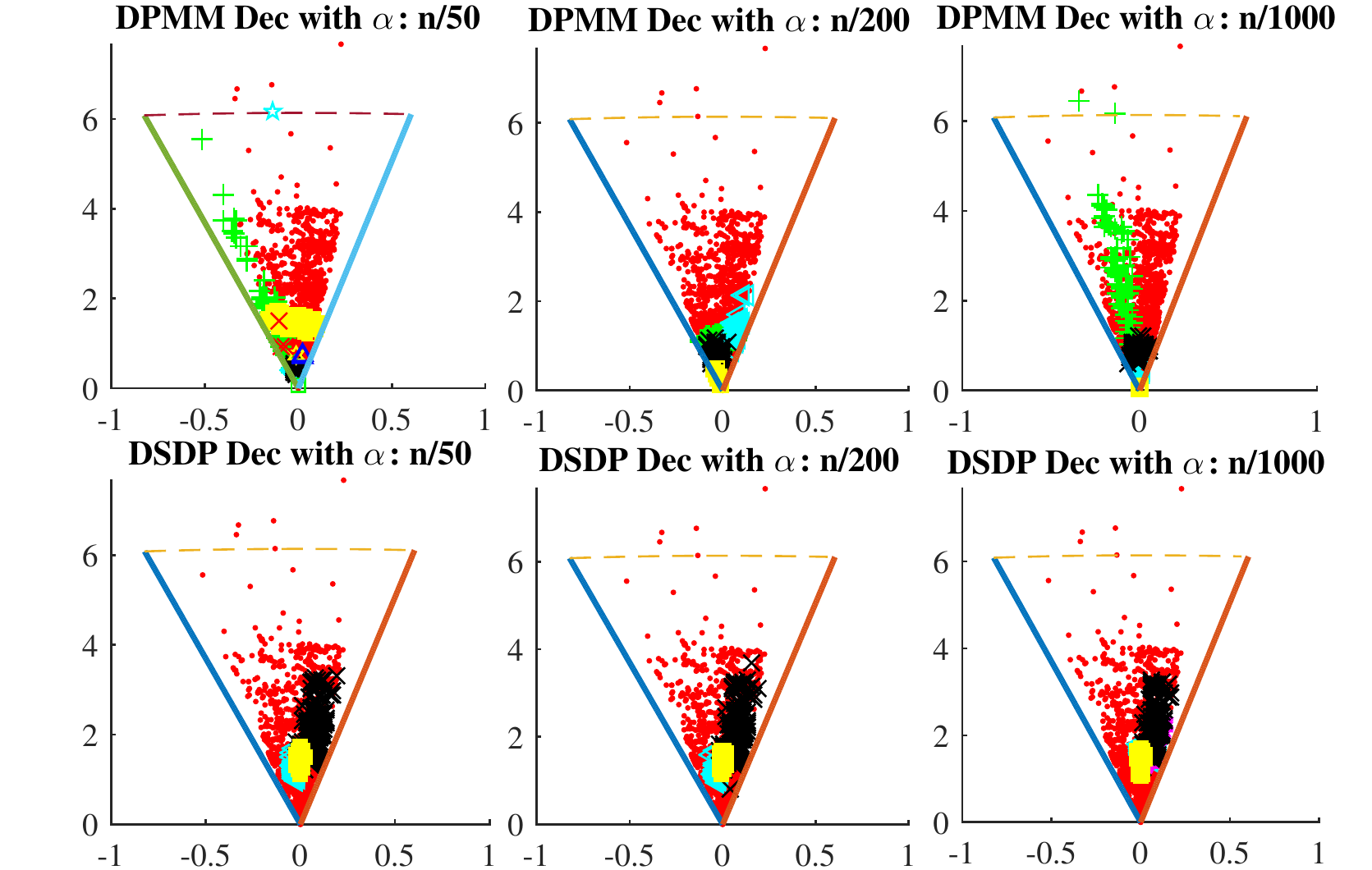}
\end{center}
\caption{ Clustering result for Galaxy dataset with different concentration parameters. The angle $\theta$ indicates "Declination", which is the coordinate in the sky similar to latitude on Earth. The x-axis and y-axis show the Cartesian coordinates of the speed (diameter) and the declination (angle) with the scale $10^{-4}\cdot v\cos{(\theta)}$ and  $10^{-4}\cdot v\sin{(\theta)}$ respectively. }
\label{Galaxy_2}
\end{figure*}

Fig.~\ref{Galaxy_2} presents the spatial clustering result of DPMM and DSDP with various $\alpha^*$. 
Spatial clusters of DPMM are varying with different $\alpha^*$, such as green, yellow and cyan clusters.
Conversely,  typical clusters of DSDP-MM in black, 
yellow and cyan, which are much robuster.

\section{Conclusion}
Contributions can be regarded in two parts for this technical report: 1) we proved the consistency for the number of components in Doubly Stochastic Dirichlet process with exponential convergence of posterior probability. We have proved this model using single and multiple clusters experiments to support the consistency proof.  2) We have introduced theoretical properties of the Doubly Stochastic Dirichlet process with the Marked SGP  prior.

\ifCLASSOPTIONcaptionsoff
  \newpage
\fi

\end{document}